\documentclass[10pt,a4paper]{article}

\usepackage[utf8]{inputenc}
\usepackage[english]{babel}

\begin{document}
\large

\newpage
\begin{center}
\LARGE
{\bf True Neutrality as a New Type of Flavour}
\end{center}
\vspace{0.1mm}
\begin{center}
{\bf Rasulkhozha S. Sharafiddinov}
\end{center}
\vspace{0.1mm}
\begin{center}
{\bf Institute of Nuclear Physics, Uzbekistan Academy of Sciences,
\\Tashkent, 100214 Ulugbek, Uzbekistan}
\end{center}
\vspace{0.1mm}

\begin{center}
{\bf Abstract}
\end{center}

A classification of leptonic currents with respect to C-operation requires the separation of elementary particles into the two classes of vector C-even and axial-vector C-odd character. 
Their nature has been created so that to each type of lepton corresponds a kind of neutrino. 
Such pairs are united in families of a different C-parity. Unlike the neutrino of a vector 
type, any C-noninvariant Dirac neutrino must have his Majorana neutrino. They constitute 
the purely neutrino families. We discuss the nature of a corresponding mechanism responsible 
for the availability in all types of axial-vector particles of a kind of flavour which 
distinguishes each of them from others by a true charge characterized by a quantum number
conserved at the interactions between the C-odd fermion and the field of emission of the corresponding types of gauge bosons. This regularity expresses the unidenticality of truly 
neutral neutrino and antineutrino, confirming that an internal symmetry of a C-noninvariant 
particle is described by an axial-vector space. Thereby, a true flavour together with the 
earlier known lepton flavour predicts the existence of leptonic strings and their birth in 
single and double beta decays as a unity of flavour and gauge symmetry laws. Such a unified 
principle explains the availability of a flavour symmetrical mode of neutrino oscillations.

\vspace{0.8cm}
\noindent
{\bf 1 Introduction}
\vspace{0.4cm}

There is no doubt that a notion about neutrino oscillation is based at first on the availability 
of solar neutrino problem but not on the absence of flavour symmetry. However, over fifty years 
ago, when the hypothesis of neutrino oscillation [1] was formulated for the first time and now, 
when a mixing of three types of neutrinos is observed [2,3], either the laboratory facts or the theoretical models are not in state to give a categorical answer to the question of whether or not an unbroken flavour symmetry exists. The answer is, as will be seen from the further, still written in matter fields [4]. 

According to old presentations about lepton nature, their interaction with the field
of emission has been described as the vector $V_{l}$ current
\begin{equation}
j_{em}^{V_{l}}=\overline{u}(p',s')[\gamma_{\mu}f_{1l}(q^{2})-
i\sigma_{\mu\lambda}q_{\lambda}f_{2l}(q^{2})]u(p,s),
\label{1}
\end{equation}
characterized by the CP-invariant components are the electric $f_{1l}$ charge and its dipole 
$f_{2l}$ moment [5]. After the opening a violation of parity [6] conservation concerning P, the possibility of the availability has become clear between the lepton and the field of emission of 
one more another type of connection. This second type of interaction may be expressed in the form 
of an axial-vector $A_{l}$ current 
\begin{equation}
j_{em}^{A_{l}}=\overline{u}(p',s')\gamma_{5}[\gamma_{\mu}
g_{1l}(q^{2})-i\sigma_{\mu\lambda}q_{\lambda}g_{2l}(q^{2})]u(p,s).
\label{2}
\end{equation}
Here as well as in (\ref{1}) $\sigma_{\mu\lambda}=[\gamma_{\mu},\gamma_{\lambda}]/2,$ $q=p-p',$ $p(s)$ and $p'(s')$ denote the four-momenta (helicities) of incoming and outgoing particles.

In quantum electrodynamics, it is usually accepted that the density of the photon leptonic 
current has the form
\begin{equation}
j_{em}=j_{em}^{V_{l}}+j_{em}^{A_{l}}.
\label{3}
\end{equation}

It includes the CP-invariant classical $g_{1l}$ anapole [7,8] and the CP-noninvariant electric $g_{2l}$ dipole [9,10] having a different T-symmetricality with the same P-antisymmetricality. 
Such a definition, however, leads to difficulties. In fact, it follows from considerations of symmetry that each C-even or C-odd dipole must arise as a result of a kind of charge [11]. 

In these circumstances, the processes on the nuclear targets may relate the mass $m_{l}$ to 
electric $f_{1l}$ charge and dipole $f_{2l}$ moment of any lepton as a unity of interaction structural parts. This connection in the low energy limit $(q^{2}\rightarrow 0)$ is reduced 
to one [12-15] of highly important consequences of the ideas of flavour symmetry: 
\begin{equation}
f_{1l}(0)-2m_{l}f_{2l}(0)=0.
\label{4}
\end{equation}

To express these ideas more clearly, one must refer to the case when to the process responds 
only the current axial-vector component. At such a situation, an equality of the interaction 
cross sections with the field of emission of both types of axial-vector $A_{l}$ currents may 
serve as a certain indication [16,17] to an explicit connection
\begin{equation}
g_{1l}(0)-m_{l}g_{2l}(0)=0.
\label{5}
\end{equation}

In the presence of all parts of the interaction (\ref{3}), the corresponding cross section for 
the process with polarized leptons includes not only self interference components $f_{il}^{2}$ 
and $g_{il}^{2},$ but also the contributions $\lambda_{c}sf_{1l}g_{1l}$ and $sf_{2l}g_{2l}$ of 
the mixed interference [18-20] between the two currents of the vector and axial-vector character. 
Among them $\lambda_{c}sf_{1l}g_{1l}$ for the particle $(\lambda_{c}=+1)$ and the antiparticle $(\lambda_{c}=-1)$ are different. A given circumstance expresses the C-antisymmetry of CP-even anapole. In contrast to this, the absence of $\lambda_{c}$ reflects, in the case of 
$f_{2l}g_{2l},$ the fact that $g_{2l}$ was accepted in (\ref{2}) as the usual 
C-invariant dipole. 

Thus, the mixed interference parts of the interaction cross section would seem to say  
in favor between $f_{il}(0)$ and $g_{il}(0)$ of a flavour symmetrical connection. Such 
a dependence, however, does not exist at all. The point is that the availability of the 
multipliers $\lambda_{c}$ and $s$ violates parity conservation concerning C and P as well 
as the above-noted equality of the interaction cross section with the field of emission
of leptonic current structural components. This in turn implies the violation of lepton 
number conservation law, confirming that we cannot exclude the validity of each of 
(\ref{4}) and (\ref{5}) only for particles of the defined type of lepton.

These facts give the right to interpret the classical $g_{1l}$ anapole as the C-odd electric charge [13,17]. Therefore, if it turns out that $f_{il}$ and $g_{il}$ constitute symbolically the vector $V_{l}$ and axial-vector $A_{l}$ components of the same Dirac $(i=1)$ or Pauli $(i=2)$ part of leptonic current, from the point of view of (\ref{4}) and (\ref{5}), it should be expected that
the dipole moments $f_{2l}$ and $g_{2l}$ correspond in nature to charges [13,17] of vector 
$f_{1l}$ and axial-vector $g_{1l}$ types.

So, we have learned that any C-invariant or C-noninvariant charge may not serve as the 
source of the two types of the dipole moments of a different C-invariance. In other words, 
C-noninvariance of the anapole charge expresses the C-oddity as well as the CP-parity 
of an electric dipole. At the same time, T-noninvariance itself of an electric dipole 
is explained by the violation of T-parity conservation of the anapole charge even at 
the absence of CPT-symmetry of all types of $A_{l}$ currents. 

Many authors state that a violation of CPT-symmetry is incompatible with the Lorentz 
invariance [21]. If we accept this idea, the absence of Lorentz symmetry would lead 
us to an extension [22] of the standard model [23-25]. 

Another characteristic moment is that regardless of whether [21] or not [26] a connection
between the CPT and the Lorentz invariance exists, each neutrino [17,27] may not be
simultaneously both a CP-even Dirac fermion and a CP-odd Majorana one [28]. At the same 
time, any type of charge cannot interact with all of the existing types of fields.

Therefore, the nature itself unites all elementary particles in families of a different 
C-invariance. To one of them apply the C-invariant leptons $(l^{V}\neq \overline{l}^{V})$ 
and their Dirac neutrinos $(\nu_{l}^{V}\neq {\bar \nu}_{l}^{V})$ with vector $V_{l}$ currents.
In these particles, the axial-vector interactions are absent. Among the fermions of another 
family one can meet the C-odd Majorana $(\nu_{M}^{A}={\bar \nu}_{M}^{A})$ neutrinos. They have 
their C-noninvariant Dirac $(\nu_{l}^{A}={\bar \nu}_{l}^{A})$ neutrinos [27] corresponding in 
united families to the kind of truly neutral leptons $(l^{A}=\overline{l}^{A})$ of 
axial-vector $A_{l}$ currents. These particles do not possess the vector interactions.
Such a separation of fermions into the two classes of vector $(V)$ and axial-vector 
$(A)$ types [13] leads us to a unified principle that
\begin{equation}
l^{V}=e^{V}, \mu^{V}, \tau^{V}, ...\rightarrow \nu_{l}^{V}=
\nu_{e}^{V}, \nu_{\mu}^{V}, \nu_{\tau}^{V}, ...,
\label{6}
\end{equation}
\begin{equation}
l^{A}=e^{A}, \mu^{A}, \tau^{A}, ...\rightarrow \nu_{l}^{A}=
\nu_{e}^{A}, \nu_{\mu}^{A}, \nu_{\tau}^{A}, ...,
\label{7}
\end{equation}
\begin{equation}
\nu_{D}^{A}=\nu_{e}^{A}, \nu_{\mu}^{A},
\nu_{\tau}^{A}, ...\rightarrow \nu_{M}^{A}=\nu_{1}^{A},
\nu_{2}^{A}, \nu_{3}^{A}, ....
\label{8}
\end{equation}

The feature of a presentation (\ref{6}) is the idea that each type of
vector lepton testifies in favor of the existence of a kind of C-invariant
neutrino. These pairs can constitute the leptonic families of a vector nature.
We must, therefore, define [29] their family structure in the form 
\begin{equation}
\pmatrix{\nu_{e}^{V}\cr e^{V}}_{L},
(\nu_{e}^{V}, \, \, \, \, e^{V})_{R}, \, \, \, \,
\pmatrix{\nu_{\mu}^{V}\cr \mu^{V}}_{L},
(\nu_{\mu}^{V}, \, \, \, \, \mu^{V})_{R}, \, \, \, \,
\pmatrix{\nu_{\tau}^{V}\cr \tau^{V}}_{L},
(\nu_{\tau}^{V}, \, \, \, \, \tau^{V})_{R}, ...,
\label{9}
\end{equation}
\begin{equation}
\pmatrix{{\bar \nu_{e}^{V}}\cr {\bar e^{V}}}_{R},
({\bar \nu_{e}^{V}}, \, \, \, \, {\bar e^{V}})_{L}, \, \, \, \,
\pmatrix{{\bar \nu_{\mu}^{V}}\cr {\bar \mu^{V}}}_{R},
({\bar \nu_{\mu}^{V}}, \, \, \, \, {\bar \mu^{V}})_{L}, \, \, \, \,
\pmatrix{{\bar \nu_{\tau}^{V}}\cr {\bar \tau^{V}}}_{R},
({\bar \nu_{\tau}^{V}}, \, \, \, \, {\bar \tau^{V}})_{L}, ....
\label{10}
\end{equation}

Any family possesses an individual flavour [30,31]. This gives the right
to characterize each particle by the three
$(l^{V}=e^{V},$ $\mu^{V},$ $\tau^{V})$ lepton flavours:
\begin{equation}
L_{l^{V}}=\left\{
{\begin{array}{l}
{+1\quad \mbox{for}\quad l^{V}_{L}, \, \, \, \, \, l^{V}_{R}, \, \, \, \,
\nu_{lL}^{V}, \, \, \, \, \nu_{lR}^{V},}\\
{-1\quad \mbox{for}\quad \overline{l}^{V}_{R}, \, \, \, \,
\overline{l}^{V}_{L}, \, \, \, \,
{\bar \nu_{lR}^{V}}, \, \, \, \, {\bar \nu_{lL}^{V}},}\\
{\, \, \, \, \, 0\quad \mbox{for}\quad \mbox{remaining particles.}}\\
\end{array}}\right.
\label{11}
\end{equation}

In the framework of a conservation law of full lepton number
\begin{equation}
L_{e^{V}}+L_{\mu^{V}}+L_{\tau^{V}}=const
\label{12}
\end{equation}
or of all forms of lepton flavours
\begin{equation}
L_{l^{V}}=const,
\label{13}
\end{equation}
a formation [29] of the left (right)-handed individual paraparticles 
\begin{equation}
(l^{V}_{L}, \overline{l}^{V}_{R}), \, \, \, \,
(l^{V}_{R}, \overline{l}^{V}_{L}),
\label{14}
\end{equation}
\begin{equation}
(\nu_{lL}^{V}, {\bar \nu_{lR}}^{V}), \, \, \, \,
(\nu_{lR}^{V}, {\bar \nu_{lL}}^{V})
\label{15}
\end{equation}
at the interaction of C-invariant leptons or neutrinos with virtual photons of a vector nature $(\gamma^{V})$ is fully possible. For example, in the elastic processes of the scattering
on a nucleus Coulomb $(Z)$ charge
\begin{equation}
l^{V}_{L,R}(\overline{l}^{V}_{R,L})+A(Z)\stackrel{\gamma^{V}} {\rightarrow}
l^{V}(\overline{l}^{V})+A(Z),
\label{16}
\end{equation}
\begin{equation}
\nu_{lL,R}^{V}({\bar \nu_{lR,L}^{V}})+A(Z)\stackrel{\gamma^{V}} {\rightarrow}
\nu_{l}^{V}({\bar \nu_{l}^{V}})+A(Z).
\label{17}
\end{equation}

Any difermion here can also explain the conservation of summed size of the C-even electric 
charge $e_{l^{V}}=f_{1l^{V}}(0).$ 

The presentation (\ref{7}) expresses the regularity that each of axial-vector
leptons has his truly neutral neutrino. They constitute the united families
of C-noninvariant fermions of a Dirac nature. Thereby, such pairs
establish their family structure:
\begin{equation}
\pmatrix{\nu_{e}^{A}\cr e^{A}}_{L},
(\nu_{e}^{A}, \, \, \, \, e^{A})_{R}, \, \, \, \,
\pmatrix{\nu_{\mu}^{A}\cr \mu^{A}}_{L},
(\nu_{\mu}^{A}, \, \, \, \, \mu^{A})_{R}, \, \, \, \,
\pmatrix{\nu_{\tau}^{A}\cr \tau^{A}}_{L},
(\nu_{\tau}^{A}, \, \, \, \, \tau^{A})_{R}, ...,
\label{18}
\end{equation}
\begin{equation}
\pmatrix{{\bar \nu_{e}^{A}}\cr {\bar e^{A}}}_{R},
({\bar \nu_{e}^{A}}, \, \, \, \, {\bar e^{A}})_{L}, \, \, \, \,
\pmatrix{{\bar \nu_{\mu}^{A}}\cr {\bar \mu^{A}}}_{R},
({\bar \nu_{\mu}^{A}}, \, \, \, \, {\bar \mu^{A}})_{L}, \, \, \, \,
\pmatrix{{\bar \nu_{\tau}^{A}}\cr {\bar \tau^{A}}}_{R},
({\bar \nu_{\tau}^{A}}, \, \, \, \, {\bar \tau^{A}})_{L}, ....
\label{19}
\end{equation}

A highly characteristic feature of this picture is the fact that a formation
of any of individual parafermions
\begin{equation}
(l^{A}_{L}, \overline{l}^{A}_{R}), \, \, \, \,
(l^{A}_{R}, \overline{l}^{A}_{L}),
\label{20}
\end{equation}
\begin{equation}
(\nu_{lL}^{A}, {\bar \nu_{lR}}^{A}), \, \, \, \,
(\nu_{lR}^{A}, {\bar \nu_{lL}}^{A})
\label{21}
\end{equation}
is responsible for conservation [17] of the summed size of the C-noninvariant electric charge $e_{l^{A}}=g_{1l^{A}}(0),$ for example, in the processes on nuclear targets
\begin{equation}
l^{A}_{L,R}(\overline{l}^{A}_{R,L})+A(Z)\stackrel{\gamma^{A}} {\rightarrow}
l^{A}(\overline{l}^{A})+A(Z),
\label{22}
\end{equation}
\begin{equation}
\nu_{lL,R}^{A}({\bar \nu_{lR,L}^{A}})+A(Z)\stackrel{\gamma^{A}} {\rightarrow}
\nu_{l}^{A}({\bar \nu_{l}^{A}})+A(Z),
\label{23}
\end{equation}
where the incoming fluxes consist of C-odd leptons or of their Dirac neutrinos. They interact 
with virtual axial-vector photons $(\gamma^{A}),$ the existence of which follows, as we will 
see below, from the local gauge invariance of the discussed types of fields.

Another important consequence implied from (\ref{8}) is that to each type of
C-noninvariant Dirac neutrino corresponds a kind of truly neutral Majorana
neutrino. The similar pairs constitute the purely neutrino families of
axial-vector fermions of a different nature. This gives the possibility
to define their family structure [27] in the following form:
\begin{equation}
\pmatrix{\nu_{e}^{A}\cr \nu_{1}^{A}}_{L},
(\nu_{e}^{A}, \, \, \, \, \nu_{1}^{A})_{R}, \, \, \, \,
\pmatrix{\nu_{\mu}^{A}\cr \nu_{2}^{A}}_{L},
(\nu_{\mu}^{A}, \, \, \, \, \nu_{2}^{A})_{R}, \, \, \, \,
\pmatrix{\nu_{\tau}^{A}\cr \nu_{3}^{A}}_{L},
(\nu_{\tau}^{A}, \, \, \, \, \nu_{3}^{A})_{R}, ...,
\label{24}
\end{equation}
\begin{equation}
\pmatrix{{\bar \nu_{e}^{A}}\cr {\bar \nu_{1}^{A}}}_{R},
({\bar \nu_{e}^{A}}, \, \, \, \, {\bar \nu_{1}^{A}})_{L}, \, \, \, \,
\pmatrix{{\bar \nu_{\mu}^{A}}\cr {\bar \nu_{2}^{A}}}_{R},
({\bar \nu_{\mu}^{A}}, \, \, \, \, {\bar \nu_{2}^{A}})_{L}, \, \, \, \,
\pmatrix{{\bar \nu_{\tau}^{A}}\cr {\bar \nu_{3}^{A}}}_{R},
({\bar \nu_{\tau}^{A}}, \, \, \, \, {\bar \nu_{3}^{A}})_{L}, ....
\label{25}
\end{equation}

Such a picture is based logically on the idea of a coexistence law of truly neutral neutrinos 
of Dirac and Majorana types [27]. Thereby, it predicts the conservation of the C-noninvariant 
axial-vector electric charge [17], for example, in the elastic scattering on a spinless nucleus
\begin{equation}
\nu_{ML,R}^{A}({\bar \nu_{MR,L}^{A}})+A(Z)\stackrel{\gamma^{A}} {\rightarrow}
\nu_{M}^{A}({\bar \nu_{M}^{A}})+A(Z)
\label{26}
\end{equation}
as a unification of left (right)-handed C-odd fermions in individual difermions
\begin{equation}
(\nu_{ML}^{A}, {\bar \nu_{MR}}^{A}), \, \, \, \,
(\nu_{MR}^{A}, {\bar \nu_{ML}}^{A}).
\label{27}
\end{equation}

On the other hand, as was noted in the work [13], any of parafermions
(\ref{14}) and (\ref{15}) can be formed only in the case where flavour
symmetry establishes the interaction P-symmetrical picture, although
P-symmetry of fermion is basically violated at the expense of its self
inertial mass [27]. If we here take into account that as well as in the
systems (\ref{14}) and (\ref{15}), the existence of a hard P-symmetrical
connection between the two particles of each of paraparticles (\ref{20}),
(\ref{21}) and (\ref{27}) is by no means excluded naturally [17], there 
arises a question of whether the hypothesis about unidenticality of truly 
neutral neutrino and antineutrino is not strictly nonverisimilar even at 
the violation of CPT-symmetry of a CP-even axial-vector fermion itself.

Our purpose in a given work is to formulate a principle, according to which, any of families (\ref{18}), (\ref{19}), (\ref{24}) and (\ref{25}) distinguishes from others by a true charge characterized by a quantum number conserved in the processes with axial-vector leptons and 
neutrinos analogously to the fact that lepton flavours or full lepton number are conserved 
at the interactions between the vector fermion and the field of emission of the corresponding 
types of gauge bosons. The true charges of all C-antisymmetrical leptons for both types of 
C-noninvariant neutrinos and antineutrinos are different and that, consequently, lead to 
the kind of truly strict selection rules.

\vspace{0.8cm}
\noindent
{\bf 2 From the earlier CP-symmetry to the new flavour}
\vspace{0.4cm}

A truly neutrality of neutrinos of a Majorana nature implies the absence in such fermions of 
a vector C-even electric charge and of each lepton flavour. But, as we have already mentioned 
above, unlike the earlier picture of the united Coulomb interactions with leptonic currents, 
their classification with respect to C-operation assumed the existence for all C-noninvariant leptons and neutrinos of a kind of axial-vector C-odd electric charge [17].

It is important that by this reason in the interaction process with 
nuclei of the two types of massive Dirac neutrinos of the vector $V$
and axial-vector $A$ nature
\begin{equation}
\nu(\nu_{lL,R}^{V}, \nu_{lL,R}^{A})+
A(Z)\stackrel{\gamma^{V}, \gamma^{A}} {\longrightarrow}
\nu(\nu_{l}^{V}, \nu_{l}^{A})+A(Z),
\label{28}
\end{equation}
\begin{equation}
{\bar \nu}({\bar \nu_{lR,L}}^{V}, {\bar \nu_{lR,L}}^{A})+
A(Z)\stackrel{\gamma^{V}, \gamma^{A}} {\longrightarrow}
{\bar \nu}({\bar \nu_{l}}^{V}, {\bar \nu_{l}}^{A})+A(Z)
\label{29}
\end{equation}
there arise not only parafermions (\ref{15}) and (\ref{21}), but also the
united systems of particles of a different C-invariance
\begin{equation}
(\nu^{V}_{lL}, {\bar \nu}^{A}_{lR}), \, \, \, \,
(\nu^{V}_{lR}, {\bar \nu}^{A}_{lL}),
\label{30}
\end{equation}
\begin{equation}
(\nu^{A}_{lL}, {\bar \nu}^{V}_{lR}), \, \, \, \,
(\nu^{A}_{lR}, {\bar \nu}^{V}_{lL}).
\label{31}
\end{equation}

The existence of any of paraneutrinos (\ref{30}) and (\ref{31}) is, as was mentioned in [13], incompatible with the parity concerning C and P as well as with the lepton number 
conservation law.

These violations would seem to say that either among the incoming particles the 
same C-even or C-odd neutrinos are present or the hypothesis about unification of 
fermions in families of a different C-parity is not valid. On the other hand, as follows 
from simple reasoning, each type of an asymmetry [20] arising at the expense of interference 
between the interactions of the symmetrical and antisymmetrical character concerning C or P, 
can be explained by a formation of a unified system of the two particles of the vector and 
axial-vector nature [32]. At the same time, nonconsevation of both C and P at the interaction 
of neutrinos of the most diverse currents with the field of emission is practically not 
excluded [33]. At our sight, this becomes possible owing to the fact that among the 
investigated and the used particles sometimes one can find as well as other types 
from the same elementary objects.

But here we have learned that an appearance in the nucleus charge field
of any of paraparticles (\ref{30}) and (\ref{31}) explains once more the
absence for all types of truly neutral leptons and neutrinos of each of the
earlier known lepton flavours $(L_{l^{V}})$ and the availability in them of
a kind of flavour $(T_{l^{A}})$ which can be called a true charge. Then it
is possible, for example, the interaction with the field of emission of 
CP-invariant parafermions such as (\ref{30}) and (\ref{31}) is not 
forbidden by a conservation law
\begin{equation}
L_{\nu^{V}_{l}}+T_{{\bar \nu}^{A}_{l}}=const, \, \, \, \,
T_{\nu^{A}_{l}}+L_{{\bar \nu}^{V}_{l}}=const.
\label{32}
\end{equation}

According to these selection rules, any of difermions (\ref{30}) and (\ref{31})
may serve as the strong argument in favor of quantum numbers
\begin{equation}
T_{\nu^{A}_{l}}=+1, \, \, \, \, T_{{\bar \nu}^{A}_{l}}=-1.
\label{33}
\end{equation}

Furthermore, each of paraneutrinos (\ref{15}) and (\ref{27}) there exists only in the presence 
of a unified P-symmetrical force between its structural particles. Such a connection appears
as a consequence of a kind of regularity of nature of elastic scattering of vector Dirac 
and axial-vector Majorana neutrinos on nuclei
\begin{equation}
\nu(\nu_{lL,R}^{V}, \nu_{ML,R}^{A})+
A(Z)\stackrel{\gamma^{V}, \gamma^{A}} {\longrightarrow}
\nu(\nu_{l}^{V}, \nu_{M}^{A})+A(Z),
\label{34}
\end{equation}
\begin{equation}
{\bar \nu}({\bar \nu_{lR,L}}^{V}, {\bar \nu_{MR,L}}^{A})+
A(Z)\stackrel{\gamma^{V}, \gamma^{A}} {\longrightarrow}
{\bar \nu}({\bar \nu_{l}}^{V}, {\bar \nu_{M}}^{A})+A(Z).
\label{35}
\end{equation}

If this is so then from the point of view of any type of flavour $L_{l^{V}}$ or $T_{l^{A}},$
it should be expected that a formation of parafermions
\begin{equation}
(\nu^{V}_{lL}, {\bar \nu}^{A}_{MR}), \, \, \, \,
(\nu^{V}_{lR}, {\bar \nu}^{A}_{ML}),
\label{36}
\end{equation}
\begin{equation}
(\nu^{A}_{ML}, {\bar \nu}^{V}_{lR}), \, \, \, \,
(\nu^{A}_{MR}, {\bar \nu}^{V}_{lL})
\label{37}
\end{equation}
from the neutrinos of a different C-invariance corresponds in united processes (\ref{34}) 
and (\ref{35}) to the existence of additive selection rules
\begin{equation}
L_{\nu^{V}_{l}}+T_{{\bar \nu}^{A}_{M}}=const, \, \, \, \,
T_{\nu^{A}_{M}}+L_{{\bar \nu}^{V}_{l}}=const.
\label{38}
\end{equation}

The latter together with ideas of each of paraparticles (\ref{36}) and (\ref{37}) predicts the 
size of a true charge for a Majorana neutrino
\begin{equation}
T_{\nu^{A}_{M}}=+1, \, \, \, \, T_{{\bar \nu}^{A}_{M}}=-1.
\label{39}
\end{equation}

Therefore, It is relevant to emphasize once more the legality [17], for example, of interactions 
of the two types of axial-vector neutrinos of the Dirac and Majorana nature with the same 
field of emission 
\begin{equation}
\nu(\nu_{lL,R}^{A}, \nu_{ML,R}^{A})+A(Z)\stackrel{\gamma^{A}} {\rightarrow}
\nu(\nu_{l}^{A}, \nu_{M}^{A})+A(Z),
\label{40}
\end{equation}
\begin{equation}
{\bar \nu}({\bar \nu_{lR,L}}^{A}, {\bar \nu_{MR,L}}^{A})+
A(Z)\stackrel{\gamma^{A}} {\rightarrow}
{\bar \nu}({\bar \nu_{l}}^{A}, {\bar \nu_{M}}^{A})+A(Z).
\label{41}
\end{equation}

They can exist only in the case where the appearance of difermions (\ref{21}) and (\ref{27}) 
as well as of C-invariant paraneutrinos of the same P-parity
\begin{equation}
(\nu^{A}_{lL}, {\bar \nu}^{A}_{MR}), \, \, \, \,
(\nu^{A}_{lR}, {\bar \nu}^{A}_{ML}),
\label{42}
\end{equation}
\begin{equation}
(\nu^{A}_{ML}, {\bar \nu}^{A}_{lR}), \, \, \, \,
(\nu^{A}_{MR}, {\bar \nu}^{A}_{lL})
\label{43}
\end{equation}
from the C-odd fermions does not contradict the conditions
\begin{equation}
T_{\nu^{A}_{l}}+T_{{\bar \nu}^{A}_{M}}=const, \, \, \, \,
T_{\nu^{A}_{M}}+T_{{\bar \nu}^{A}_{l}}=const.
\label{44}
\end{equation}

Finally, insofar as a question about quantum numbers $T_{l^{A}}$ and
$T_{\overline{l}^{A}},$ we will start from the elastic scattering of leptons
$(l=e,$ $\mu,$ $\tau, ...)$ of a different C-parity on the target nuclei
\begin{equation}
l(l^{V}_{L,R}, l^{A}_{L,R})+
A(Z)\stackrel{\gamma^{V}, \gamma^{A}} {\longrightarrow}
l(l^{V}, l^{A})+A(Z),
\label{45}
\end{equation}
\begin{equation}
\overline{l}(\overline{l}^{V}_{R,L}, \overline{l}^{A}_{R,L})+
A(Z)\stackrel{\gamma^{V}, \gamma^{A}} {\longrightarrow}
\overline{l}(\overline{l}^{V}, \overline{l}^{A})+A(Z).
\label{46}
\end{equation}

It originates only if unlike the individual dileptons (\ref{14}) and (\ref{20}),
the united paraparticles
\begin{equation}
(l^{V}_{L}, \overline{l}^{A}_{R}), \, \, \, \,
(l^{V}_{R}, \overline{l}^{A}_{L}),
\label{47}
\end{equation}
\begin{equation}
(l^{A}_{L}, \overline{l}^{V}_{R}), \, \, \, \,
(l^{A}_{R}, \overline{l}^{V}_{L}),
\label{48}
\end{equation}
appear jointly with the selection rules
\begin{equation}
L_{l^{V}}+T_{\overline{l}^{A}}=const, \, \, \, \,
T_{l^{A}}+L_{\overline{l}^{V}}=const.
\label{49}
\end{equation}

Comparing their with that constitute difermions (\ref{14}) and (\ref{20}),
but having in mind the paraleptons (\ref{47}) and (\ref{48}), one can
found from (\ref{49}) that
\begin{equation}
T_{l^{A}}=+1, \, \, \, \, T_{\overline{l}^{A}}=-1.
\label{50}
\end{equation}

To verify the signs in sizes of (\ref{33}), (\ref{39}) and (\ref{50}) one must
apply to the processes with neutrinos of the most diverse types. A beautiful
example is the muon decay of a vector nature
\begin{equation}
\mu^{V}_{L,R}\stackrel{\gamma^{V}} {\rightarrow}
e^{V}_{L,R}{\bar \nu_{eR,L}^{V}}\nu_{\mu L,R}^{V}, \, \, \, \,
{\bar \mu^{V}_{R,L}}\stackrel{\gamma^{V}} {\rightarrow}
{\bar e^{V}_{R,L}}\nu_{eL,R}^{V}{\bar \nu_{\mu R,L}^{V}}.
\label{51}
\end{equation}

These transitions say in favor of the C-even electric charge conservation
\begin{equation}
e_{\mu^{V}}=e_{e^{V}}+e_{{\bar \nu_{e}^{V}}}+e_{\nu_{\mu}^{V}},
\label{52}
\end{equation}
\begin{equation}
e_{{\bar \mu^{V}}}=e_{{\bar e^{V}}}+e_{\nu_{e}^{V}}+e_{{\bar \nu_{\mu}^{V}}},
\label{53}
\end{equation}
confirming that conditions
\begin{equation}
\Delta L_{\mu^{V}}=
L_{\mu^{V}}-L_{\nu_{\mu}^{V}}=L_{e^{V}}+L_{{\bar \nu_{e}}^{V}},
\label{54}
\end{equation}
\begin{equation}
\Delta L_{{\bar \mu^{V}}}=
L_{{\bar \mu^{V}}}-L_{{\bar \nu_{\mu}^{V}}}=L_{{\bar e}^{V}}+L_{\nu_{e}^{V}}
\label{55}
\end{equation}
correspond in (\ref{51}) to a formation of vector difermions
\begin{equation}
(e^{V}_{L}, {\bar \nu_{eR}}^{V}), \, \, \, \,
(e^{V}_{R}, {\bar \nu_{eL}}^{V}),
\label{56}
\end{equation}
\begin{equation}
({\bar e_{R}}^{V}, \nu_{eL}^{V}), \, \, \, \,
({\bar e_{L}}^{V}, \nu_{eR}^{V}).
\label{57}
\end{equation}

To a similar implication one can lead starting from the $\mu^{A}$-decay
\begin{equation}
\mu^{A}_{L,R}\stackrel{\gamma^{A}} {\rightarrow}
e^{A}_{L,R}{\bar \nu_{eR,L}^{A}}\nu_{\mu L,R}^{A}, \, \, \, \,
{\bar \mu^{A}_{R,L}}\stackrel{\gamma^{A}} {\rightarrow}
{\bar e^{A}_{R,L}}\nu_{eL,R}^{A}{\bar \nu_{\mu R,L}^{A}}.
\label{58}
\end{equation}
Here the appearance of axial-vector parafermions
\begin{equation}
(e^{A}_{L}, {\bar \nu_{eR}}^{A}), \, \, \, \,
(e^{A}_{R}, {\bar \nu_{eL}}^{A}),
\label{59}
\end{equation}
\begin{equation}
({\bar e_{R}}^{A}, \nu_{eL}^{A}), \, \, \, \,
({\bar e_{L}}^{A}, \nu_{eR}^{A})
\label{60}
\end{equation}
responds to conservation of both true number
\begin{equation}
\Delta T_{\mu^{A}}=
T_{\mu^{A}}-T_{\nu_{\mu}^{A}}=T_{e^{A}}+T_{{\bar \nu_{e}}^{A}},
\label{61}
\end{equation}
\begin{equation}
\Delta T_{{\bar \mu^{A}}}=
T_{{\bar \mu^{A}}}-T_{{\bar \nu_{\mu}^{A}}}=T_{{\bar e}^{A}}+T_{\nu_{e}^{A}},
\label{62}
\end{equation}
and C-odd electric charge
\begin{equation}
e_{\mu^{A}}=e_{e^{A}}+e_{{\bar \nu_{e}^{A}}}+e_{\nu_{\mu}^{A}},
\label{63}
\end{equation}
\begin{equation}
e_{{\bar \mu^{A}}}=e_{{\bar e^{A}}}+e_{\nu_{e}^{A}}+e_{{\bar \nu_{\mu}^{A}}}.
\label{64}
\end{equation}

Such a picture, however, loses the thought in decays
\begin{equation}
\mu^{V}_{L,R}\stackrel{\gamma^{V}, \gamma^{A}} {\longrightarrow}
e^{A}_{L,R}{\bar \nu_{eR,L}^{A}}\nu_{\mu L,R}^{V}, \, \, \, \,
{\bar \mu^{V}_{R,L}}\stackrel{\gamma^{V}, \gamma^{A}} {\longrightarrow}
{\bar e^{A}_{R,L}}\nu_{eL,R}^{A}{\bar \nu_{\mu R,L}^{V}},
\label{65}
\end{equation}
where the paraparticles (\ref{59}) and (\ref{60}) appear owing
to the selection rules
\begin{equation}
\Delta L_{\mu^{V}}=
L_{\mu^{V}}-L_{\nu_{\mu}^{V}}=T_{e^{A}}+T_{{\bar \nu_{e}}^{A}},
\label{66}
\end{equation}
\begin{equation}
\Delta L_{{\bar \mu^{V}}}=
L_{{\bar \mu^{V}}}-L_{{\bar \nu_{\mu}^{V}}}=T_{{\bar e}^{A}}+T_{\nu_{e}^{A}}.
\label{67}
\end{equation}

They satisfy the inequalities
\begin{equation}
e_{\mu^{V}}\neq e_{e^{A}}+e_{{\bar \nu_{e}^{A}}}+e_{\nu_{\mu}^{V}},
\label{68}
\end{equation}
\begin{equation}
e_{{\bar \mu^{V}}}\neq e_{{\bar e^{A}}}+e_{\nu_{e}^{A}}+e_{{\bar \nu_{\mu}^{V}}},
\label{69}
\end{equation}
which follow from the fact that a distinction of the truly neutral fermion and its antiparticle
\begin{equation}
l^{A}\neq \overline{l}^{A}, \, \, \, \,
\nu_{l}^{A}\neq {\bar \nu_{l}}^{A}, \, \, \, \,
\nu_{M}^{A}\neq {\bar \nu_{M}}^{A}
\label{70}
\end{equation}
cannot take place with their respect to C-operation, and is the consequence of an exact 
CP-invariance of a C-odd particle itself.

Thus, although dileptons (\ref{56}) and (\ref{57}) together with the conditions
\begin{equation}
e_{\mu^{A}}\neq e_{e^{V}}+e_{{\bar \nu_{e}^{V}}}+e_{\nu_{\mu}^{A}},
\label{71}
\end{equation}
\begin{equation}
e_{{\bar \mu^{A}}}\neq e_{{\bar e^{V}}}+e_{\nu_{e}^{V}}+e_{{\bar \nu_{\mu}^{A}}}
\label{72}
\end{equation}
indicate that
\begin{equation}
\Delta T_{\mu^{A}}=
T_{\mu^{A}}-T_{\nu_{\mu}^{A}}=L_{e^{V}}+L_{{\bar \nu_{e}}^{V}},
\label{73}
\end{equation}
\begin{equation}
\Delta T_{{\bar \mu^{A}}}=
T_{{\bar \mu^{A}}}-T_{{\bar \nu_{\mu}^{T}}}=L_{{\bar e}^{V}}+L_{\nu_{e}^{V}},
\label{74}
\end{equation}
their formation in the processes
\begin{equation}
\mu^{A}_{L,R}\stackrel{\gamma^{V}, \gamma^{A}} {\longrightarrow}
e^{V}_{L,R}{\bar \nu_{eR,L}^{V}}\nu_{\mu L,R}^{A}, \, \, \, \,
{\bar \mu^{A}_{R,L}}\stackrel{\gamma^{V}, \gamma^{A}} {\longrightarrow}
{\bar e^{V}_{R,L}}\nu_{eL,R}^{V}{\bar \nu_{\mu R,L}^{A}}
\label{75}
\end{equation}
is forbidden by a charge conservation law.

Basing on the above-noted contradictions one can think that both $T_{l^{A}}$
and $T_{\overline{l}^{A}}$ in any of (\ref{33}), (\ref{39}) and (\ref{50})
must have the opposite signs. In other words, charge nonconservation in the
decays (\ref{65}) and (\ref{75}) assumed the existence of relations
among the flavours
\begin{equation}
T_{l^{A}}=-L_{l^{V}}, \, \, \, \,
T_{\overline{l}^{A}}=-L_{\overline{l}^{V}},
\label{76}
\end{equation}
\begin{equation}
T_{\nu_{l}^{A}}=-L_{\nu_{l}^{V}}, \, \, \, \,
T_{{\bar \nu_{l}^{A}}}=-L_{{\bar \nu_{l}^{V}}}.
\label{77}
\end{equation}
\begin{equation}
T_{\nu_{M}^{A}}=-L_{\nu_{l}^{V}}, \, \, \, \,
T_{{\bar \nu_{M}^{A}}}=-L_{{\bar \nu_{l}^{V}}}.
\label{78}
\end{equation}

From such a point of view, each of (\ref{66}) and (\ref{67}) can be leaded
to his truly physical form
\begin{equation}
\Delta L_{\mu^{V}}\neq T_{e^{A}}+T_{{\bar \nu_{e}}^{A}}, \, \, \, \,
\Delta L_{\mu^{V}}=+const, \, \, \, \,
T_{e^{A}}+T_{{\bar \nu_{e}}^{A}}=-const,
\label{79}
\end{equation}
\begin{equation}
\Delta L_{{\bar \mu^{V}}}\neq T_{{\bar e}^{A}}+T_{\nu_{e}^{A}}, \, \, \, \,
\Delta L_{{\bar \mu^{V}}}=-const, \, \, \, \,
T_{{\bar e}^{A}}+T_{\nu_{e}^{A}}=+const,
\label{80}
\end{equation}
and equations (\ref{73}) and (\ref{74}) in this case are replaced
by the inequalities
\begin{equation}
\Delta T_{\mu^{A}}\neq L_{e^{V}}+L_{{\bar \nu_{e}}^{V}}, \, \, \, \,
\Delta T_{\mu^{A}}=-const, \, \, \, \,
L_{e^{V}}+L_{{\bar \nu_{e}}^{V}}=+const,
\label{81}
\end{equation}
\begin{equation}
\Delta T_{{\bar \mu^{A}}}\neq L_{{\bar e}^{V}}+L_{\nu_{e}^{V}}, \, \, \, \,
\Delta T_{{\bar \mu^{A}}}=+const, \, \, \, \,
L_{{\bar e}^{V}}+L_{\nu_{e}^{V}}=-const.
\label{82}
\end{equation}

Choices of types (\ref{76})-(\ref{78}) establish in addition a flavour
symmetrical picture of elastic scattering in which the absence of parity
nonconservation connected with (\ref{32}), (\ref{38}) and (\ref{49}) is
explained by the availability of fundamentally important differences in
the charges as well as in the masses [17] of vector and axial-vector 
fermions. Without loss of generality, we may write  
\begin{equation}
e_{l^{A}}\neq e_{l^{V}}, \, \, \, \, 
e_{\overline{l}^{A}}\neq e_{\overline{l}^{V}},
\label{83}
\end{equation}
\begin{equation}
e_{\nu_{l}^{A}}\neq e_{\nu_{l}^{V}}, \, \, \, \,
e_{{\bar \nu_{l}}^{A}}\neq e_{{\bar \nu_{l}}^{V}},
\label{84}
\end{equation}
\begin{equation}
e_{\nu_{M}^{A}}\neq e_{\nu_{l}^{V}}, \, \, \, \,
e_{{\bar \nu_{M}}^{A}}\neq e_{{\bar \nu_{l}}^{V}},
\label{85}
\end{equation}
\begin{equation}
m_{l^{A}}\neq m_{l^{V}}, \, \, \, \, 
m_{\overline{l}^{A}}\neq m_{\overline{l}^{V}},
\label{86}
\end{equation}
\begin{equation}
m_{\nu_{l}^{A}}\neq m_{\nu_{l}^{V}}, \, \, \, \,
m_{{\bar \nu_{l}}^{A}}\neq m_{{\bar \nu_{l}}^{V}},
\label{87}
\end{equation}
\begin{equation}
m_{\nu_{M}^{A}}\neq m_{\nu_{l}^{V}}, \, \, \, \,
m_{{\bar \nu_{M}}^{A}}\neq m_{{\bar \nu_{l}}^{V}}.
\label{88}
\end{equation}

These consequences of flavour symmetry just and give the possibility to
understand the hypothesis about unidenticality of fermions of the vector
and axial-vector nature
\begin{equation}
l^{A}\neq l^{V}, \, \, \, \, \overline{l}^{A}\neq \overline{l}^{V},
\label{89}
\end{equation}
\begin{equation}
\nu_{l}^{A}\neq \nu_{l}^{V}, \, \, \, \,
{\bar \nu_{l}}^{A}\neq {\bar \nu_{l}}^{V},
\label{90}
\end{equation}
\begin{equation}
\nu_{M}^{A}\neq \nu_{l}^{V}, \, \, \, \,
{\bar \nu_{M}}^{A}\neq {\bar \nu_{l}}^{V}.
\label{91}
\end{equation}

So, it is seen that a true number distinguishing a C-odd particle from his
antiparticle is not of those lepton numbers, an introduction of which can be
based on the unidenticality of neutrino and antineutrino of vector currents
from the point of view of C-invariance of such fermions. At the same time,
to any of both types of quantum numbers corresponds a kind of family.
This reflects the availability of a unified regularity of their family
structure depending on flavour dynamics of particles.

Thus, our presentations about the nature of truly neutral neutrinos allow
to formulate the laws of a corresponding mechanism responsible for separation
of axial-vector fermions among the other electroweakly and strongly interacted
particles. They state that each of families (\ref{18}), (\ref{19}), (\ref{24})
and (\ref{25}) possesses his true flavour. We can, therefore, characterize any
particle by the three $(l^{A}=e^{A},$ $\mu^{A},$ $\tau^{A})$ true flavours:
\begin{equation}
T_{l^{A}}=\left\{
{\begin{array}{l}
{-1\quad \mbox{for}\quad l^{A}_{L}, \, \, \, \, \, l^{A}_{R}, \, \, \, \,
\nu_{lL}^{A}, \, \, \, \, \nu_{lR}^{A}, \, \, \, \,
\nu_{ML}^{A}, \, \, \, \, \nu_{MR}^{A},}\\
{+1\quad \mbox{for}\quad \overline{l}^{A}_{R}, \, \, \, \,
\overline{l}^{A}_{L}, \, \, \, \,
{\bar \nu_{lR}^{A}}, \, \, \, \, {\bar \nu_{lL}^{A}}, \, \, \, \,
{\bar \nu_{MR}^{A}}, \, \, \, \, {\bar \nu_{ML}^{A}},}\\
{\, \, \, \, \, 0\quad \mbox{for}\quad \mbox{remaining particles.}}\\
\end{array}}\right.
\label{92}
\end{equation}

Conservation of full true number
\begin{equation}
T_{e^{A}}+T_{\mu^{A}}+T_{\tau^{A}}=const
\label{93}
\end{equation}
or all types of true flavours
\begin{equation}
T_{l^{A}}=const
\label{94}
\end{equation}
indicates to a principle that between the two left (right)-handed fermions
from the families (\ref{18}), (\ref{19}), (\ref{24}) and (\ref{25}) there
exists a unified force. It has an important consequence for the above-noted
difermions of $A_{l}$ currents.

Of course, their appearance in the field of emission is compatible with the
ideas of each of C and P as well as with the equality of summed charge of the
interacted particles before and after the exchange by a virtual gauge boson.
Such a principle is the unified and does not depend of whether the parafermions
have a vector or an axial-vector nature. Then it is possible, for example, to
interpret the P-symmetry as a conservation law not only of flavour [12,13],
but also of any type of the electric charge.

\vspace{0.8cm}
\noindent
{\bf 3 Flavour symmetry criterion for lepton universality}
\vspace{0.4cm}

There exists a range of the structural phenomena in which the dynamical
properties of P-parity become fully crucial. An example of this may be a
decay of $\tau$-lepton of a vector nature
\begin{equation}
\tau^{V}_{L,R}\stackrel{\gamma^{V}} {\rightarrow}
e^{V}_{L,R}{\bar \nu_{eR,L}^{V}}\nu_{\tau L,R}^{V}, \, \, \, \,
{\bar \tau^{V}_{R,L}}\stackrel{\gamma^{V}} {\rightarrow}
{\bar e^{V}_{R,L}}\nu_{eL,R}^{V}{\bar \nu_{\tau R,L}^{V}}.
\label{95}
\end{equation}

As well as in (\ref{51}), each of difermions (\ref{56}) and (\ref{57}) arises
here at the conservation simultaneously of both lepton number
\begin{equation}
\Delta L_{\tau^{V}}=
L_{\tau^{V}}-L_{\nu_{\tau}^{V}}=L_{e^{V}}+L_{{\bar \nu_{e}}^{V}},
\label{96}
\end{equation}
\begin{equation}
\Delta L_{{\bar \tau^{V}}}=
L_{{\bar \tau^{V}}}-L_{{\bar \nu_{\tau}^{V}}}=L_{{\bar e}^{V}}+L_{\nu_{e}^{V}},
\label{97}
\end{equation}
and vector C-even electric charge
\begin{equation}
e_{\tau^{V}}=e_{e^{V}}+e_{{\bar \nu_{e}^{V}}}+e_{\nu_{\tau}^{V}},
\label{98}
\end{equation}
\begin{equation}
e_{{\bar \tau^{V}}}=e_{{\bar e^{V}}}+e_{\nu_{e}^{V}}+e_{{\bar \nu_{\tau}^{V}}}.
\label{99}
\end{equation}

The solutions (\ref{96})-(\ref{99}) coincide with the corresponding sizes
from (\ref{52})-(\ref{55}) and that, consequently, the connections
\begin{equation}
L_{e^{V}}=L_{\mu^{V}}=L_{\tau^{V}}, \, \, \, \,
L_{{\bar e^{V}}}=L_{{\bar \mu^{V}}}=L_{{\bar \tau^{V}}},
\label{100}
\end{equation}
\begin{equation}
L_{\nu_{e}^{V}}=L_{\nu_{\mu}^{V}}=L_{\nu_{\tau}^{V}}, \, \, \, \,
L_{{\bar \nu_{e}}^{V}}=L_{{\bar \nu_{\mu}}^{V}}=L_{{\bar \nu_{\tau}}^{V}}
\label{101}
\end{equation}
say about lepton universality of a vector interaction
\begin{equation}
e_{e^{V}}=e_{\mu^{V}}=e_{\tau^{V}}, \, \, \, \,
e_{{\bar e^{V}}}=e_{{\bar \mu^{V}}}=e_{{\bar \tau^{V}}},
\label{102}
\end{equation}
\begin{equation}
e_{\nu_{e}^{V}}=e_{\nu_{\mu}^{V}}=e_{\nu_{\tau}^{V}}, \, \, \, \,
e_{{\bar \nu_{e}}^{V}}=e_{{\bar \nu_{\mu}}^{V}}=e_{{\bar \nu_{\tau}}^{V}}.
\label{103}
\end{equation}

We remark that a formation of any of dileptons (\ref{54}) and (\ref{60}) 
in the decays of $\tau^{A}$-lepton
\begin{equation}
\tau^{A}_{L,R}\stackrel{\gamma^{A}} {\rightarrow}
e^{A}_{L,R}{\bar \nu_{eR,L}^{A}}\nu_{\tau L,R}^{A}, \, \, \, \,
{\bar \tau^{A}_{R,L}}\stackrel{\gamma^{A}} {\rightarrow}
{\bar e^{A}_{R,L}}\nu_{eL,R}^{A}{\bar \nu_{\tau R,L}^{A}}
\label{104}
\end{equation}
is not forbidden by a selection rule
\begin{equation}
\Delta T_{\tau^{A}}=
T_{\tau^{A}}-T_{\nu_{\tau}^{A}}=T_{e^{A}}+T_{{\bar \nu_{e}}^{A}},
\label{105}
\end{equation}
\begin{equation}
\Delta T_{{\bar \tau^{A}}}=
T_{{\bar \tau^{A}}}-T_{{\bar \nu_{\tau}^{A}}}=T_{{\bar e}^{A}}+T_{\nu_{e}^{A}}
\label{106}
\end{equation}
and by a conservation law of an axial-vector C-odd electric charge
\begin{equation}
e_{\tau^{A}}=e_{e^{A}}+e_{{\bar \nu_{e}^{A}}}+e_{\nu_{\tau}^{A}},
\label{107}
\end{equation}
\begin{equation}
e_{{\bar \tau^{A}}}=e_{{\bar e^{A}}}+e_{\nu_{e}^{A}}+e_{{\bar \nu_{\tau}^{A}}}.
\label{108}
\end{equation}

Their comparison with (\ref{61})-(\ref{64}) leads us once more to
an implication that
\begin{equation}
T_{e^{A}}=T_{\mu^{A}}=T_{\tau^{A}}, \, \, \, \,
T_{{\bar e^{A}}}=T_{{\bar \mu^{A}}}=T_{{\bar \tau^{A}}},
\label{109}
\end{equation}
\begin{equation}
T_{\nu_{e}^{A}}=T_{\nu_{\mu}^{A}}=T_{\nu_{\tau}^{A}}, \, \, \, \,
T_{{\bar \nu_{e}}^{A}}=T_{{\bar \nu_{\mu}}^{A}}=T_{{\bar \nu_{\tau}}^{A}},
\label{110}
\end{equation}
the existence of which can also be based simply on the lepton universality
of an axial-vector interaction
\begin{equation}
e_{e^{A}}=e_{\mu^{A}}=e_{\tau^{A}}, \, \, \, \,
e_{{\bar e^{A}}}=e_{{\bar \mu^{A}}}=e_{{\bar \tau^{A}}},
\label{111}
\end{equation}
\begin{equation}
e_{\nu_{e}^{A}}=e_{\nu_{\mu}^{A}}=e_{\nu_{\tau}^{A}}, \, \, \, \,
e_{{\bar \nu_{e}}^{A}}=e_{{\bar \nu_{\mu}}^{A}}=e_{{\bar \nu_{\tau}}^{A}}.
\label{112}
\end{equation}

So, we can conclude that each type of flavour symmetry expresses the lepton universality 
of a kind of interaction as the one of dynamical aspects of its P-invariance.

\vspace{0.8cm}
\noindent
{\bf 4 Family structure of nucleons}
\vspace{0.4cm}

Any of currents responsible for interactions of leptons and hadrons with virtual gauge 
bosons can symbolically be expressed in the form of a sum of vector and axial-vector parts 
of the same charge or dipole moment. This does not imply of course that the same neutron 
or proton must be simultaneously both a C-even and a C-odd nucleon. We have, thus, a full 
analogy to the fact that a classification of elementary particles with respect to C-operation 
admits the existence of the two types of Dirac fermions of the vector $(V)$ and axial-vector 
$(A)$ nature. If such objects are the neutrons and protons, they will constitute the nucleonic 
$(N=n,$ $p)$ families of doublets of a different C-invariance.

One group consisting of C-even nucleons of vector $V_{N}$ currents may be presented as
\begin{equation}
\pmatrix{n^{V}\cr p^{V}}_{L}, (n^{V}, \, \, \, \, p^{V})_{R},
\label{113}
\end{equation}
\begin{equation}
\pmatrix{\overline{n}^{V}\cr \overline{p}^{V}}_{R},
(\overline{n}^{V}, \, \, \, \, \overline{p}^{V})_{L}.
\label{114}
\end{equation}

The second class includes the truly neutral C-odd nucleons of axial-vector $A_{N}$ currents
\begin{equation}
\pmatrix{n^{A}\cr p^{A}}_{L}, (n^{A}, \, \, \, \, p^{A})_{R},
\label{115}
\end{equation}
\begin{equation}
\pmatrix{\overline{n}^{A}\cr \overline{p}^{A}}_{R},
(\overline{n}^{A}, \, \, \, \, \overline{p}^{A})_{L}.
\label{116}
\end{equation}

For elucidation of their ideas, it is desirable to apply
at first to the processes
\begin{equation}
n_{L,R}^{V}\stackrel{\gamma^{V}} {\rightarrow}
p_{L,R}^{V}\overline{e}_{R,L}^{V}\nu_{eL,R}^{V}, \, \, \, \,
\overline{n}_{R,L}^{V}\stackrel{\gamma^{V}} {\rightarrow}
\overline{p}_{R,L}^{V}e_{L,R}^{V}{\bar \nu_{eR,L}^{V}},
\label{117}
\end{equation}
from which it follows that vector neutron and neutrino possess the same
CP-odd electric charge
\begin{equation}
e_{n^{V}}=e_{\nu_{e}^{V}}, \, \, \, \,
e_{\overline{n}^{V}}=e_{{\bar \nu_{e}^{V}}}.
\label{118}
\end{equation}

Exactly the same one can reanalyze the decays
\begin{equation}
n_{L,R}^{A}\stackrel{\gamma^{A}} {\rightarrow}
p_{L,R}^{A}\overline{e}_{R,L}^{A}\nu_{eL,R}^{A}, \, \, \, \,
\overline{n}_{R,L}^{A}\stackrel{\gamma^{A}} {\rightarrow}
\overline{p}_{R,L}^{A}e_{L,R}^{A}{\bar \nu_{eR,L}^{A}}
\label{119}
\end{equation}
in conformity with the baryon [34] and true number conservation laws. In this case, it is 
expected that 
\begin{equation}
e_{n^{A}}=e_{\nu_{e}^{A}}, \, \, \, \,
e_{\overline{n}^{A}}=e_{{\bar \nu_{e}^{A}}},
\label{120}
\end{equation}
and thus, truly neutral neutron and neutrino do not distinguish from one another by the 
availability in them of an equal CP-even electric charge.

\vspace{0.8cm}
\noindent
{\bf 5 From the flavour symmetry to the leptonic string}
\vspace{0.4cm}

The birth of a dilepton originates in any process of $\beta$-decay by the same reason. 
Such a reason can, for example, be existence in all leptonic families of a unified 
flavour symmetrical force between the two left (right)-handed fermions of each type. 
It establishes those connections, at which there exist the left (right)-handed leptons 
in difermions comparatively for a long time without converting into the right 
(left)-handed ones, although this is not forbidden. In other words, a flavour 
symmetrical force relates the two left (right)-handed leptons in flavourless 
dileptons. They are conserved in the form of leptonic strings until an external 
action is able to separate their by a part in the particle type dependence. Therefore, 
to understand the earlier known experimental facts at the fundamental dynamical level, 
one must use a principle that a single or a double $\beta$-decay is carried out in nuclei 
without neutrinos as well as without electrons by the schemes
\begin{equation}
n_{L,R}\rightarrow
p_{L,R}+(\overline{e}_{R,L}, \nu_{eL,R}),
\label{121}
\end{equation}
\begin{equation}
\overline{n}_{R,L}\rightarrow
\overline{p}_{R,L}+(e_{L,R}, {\bar \nu_{eR,L}}),
\label{122}
\end{equation}
\begin{equation}
2n_{L,R}\rightarrow
2p_{L,R}+2(\overline{e}_{R,L}, \nu_{eL,R}),
\label{123}
\end{equation}
\begin{equation}
2\overline{n}_{R,L}\rightarrow
2\overline{p}_{R,L}+2(e_{L,R}, {\bar \nu_{eR,L}}).
\label{124}
\end{equation}

In direct experiments left (right)-handed dileptons are observed instead of electrons. 
In addition, it is necessary to take into account the transitions
\begin{equation}
(e_{L,R}, {\bar \nu_{eR,L}})\leftrightarrow
(e_{R,L}, {\bar \nu_{eL,R}}),
\label{125}
\end{equation}
\begin{equation}
(\overline{e}_{R,L}, \nu_{eL,R})\leftrightarrow
(\overline{e}_{L,R}, \nu_{eR,L}),
\label{126}
\end{equation}
because they correspond to the interconversions of nucleons of the different components.

Thus, all properties of electrons fixed in single or double $\beta$-decay one must consider as the characteristic features that refer doubtless only to a leptonic string that unites the electron and its antineutrino. Of course, any of them can also be observed in the form of a free particle. This, however, requires the creation devices of a sufficiently high sensitivity.

\vspace{0.8cm}
\noindent
{\bf 6 Coulomb fields of a different nature}
\vspace{0.4cm}

One of the most highlighted features of a classification of elementary particles with respect 
to C-operation is its notion about axial-vector photons $(\gamma^{A})$ having an electric nature. Their existence, as was mentioned earlier, may also be accepted as a consequence of invariance 
of the Dirac Lagrangian of a C-noninvariant lepton
\begin{equation}
L_{0}^{A}=\overline{\psi}_{{l}^{A}}
\gamma_{5}(i\gamma_{\mu}\partial_{\mu}-m_{{l}^{A}})\psi_{{l}^{A}}
\label{127}  
\end{equation} 
concerning the local axial-vector gauge transformation 
\begin{equation}
\psi'_{{l}^{A}}(x)=U_{A}\psi_{{l}^{A}}(x), \, \, \, \, U_{A}=e^{i\beta (x)\gamma_{5}}.
\label{128}  
\end{equation} 

In the absence of the space-time coordinate dependence, $\beta$ becomes here global phase.
At such a phase, (\ref{128}) takes the form of the global gauge invariance, and the Lagrangian $L_{0}^{A}$ is invariant concerning its action.

However, at the use of the local phase $\beta (x),$ the expected structure of the Lagrangian
encounters the appearance of an additional term
\begin{equation}
L_{0}^{A'}=
i\overline{\psi}'_{{l}^{A}}\gamma_{5}\gamma_{\mu}\partial_{\mu}\psi'_{{l}^{A}}-
m_{{l}^{A}}\overline{\psi}'_{{l}^{A}}\gamma_{5}\psi'_{{l}^{A}}=
L_{0}+\overline{\psi}_{{l}^{A}}\gamma_{\mu}
\partial_{\mu}\beta\psi_{{l}^{A}}
\label{129}  
\end{equation} 
violating its gauge invariance.

To restore this symmetry, we introduce a new axial-vector field $A_{\mu}^{A}$ which at the
fulfilment of (\ref{128}) has the following gauge transformation: 
\begin{equation}
A_{\mu}^{A'}=A_{\mu}^{A}+\frac{1}{e_{l^{A}}}\gamma_{5}\partial_{\mu}\beta.
\label{130}  
\end{equation} 

The introduced field due to his C-noninvariant nature has no interaction with a vector 
lepton but possesses with a truly neutral lepton a kind of axial-vector interaction. It 
may be expressed by the Lagrangian
\begin{equation}
L_{int}^{A}=e_{l^{A}}j_{em}^{A_{l}}A_{\mu}^{A}=
e_{l^{A}}\overline{\psi}_{{l}^{A}}\gamma_{5}\gamma_{\mu}\psi_{{l}^{A}}A_{\mu}^{A}.
\label{131}  
\end{equation} 

Making the explicit gauge transformations, we found
\begin{equation}
L_{int}^{A'}=e_{l^{A}}\overline{\psi}'_{{l}^{A}}
\gamma_{5}\gamma_{\mu}\psi'_{{l}^{A}}A_{\mu}^{A'}=
L_{int}^{A}-\overline{\psi}_{{l}^{A}}\gamma_{\mu}
\partial_{\mu}\beta\psi_{{l}^{A}}.
\label{132}  
\end{equation} 

This together with (\ref{127}) convinces us here that 
\begin{equation}
L^{A}=\overline{\psi}_{{l}^{A}}
\gamma_{5}(i\gamma_{\mu}\partial_{\mu}-m_{{l}^{A}})\psi_{{l}^{A}}+
e_{l^{A}}\overline{\psi}_{{l}^{A}}\gamma_{5}\gamma_{\mu}\psi_{{l}^{A}}A_{\mu}^{A}.
\label{133}  
\end{equation}

The presence of mass of field $A_{\mu}^{A}$ in the Lagrangian would imply its noninvariance concerning the chosen gauge transformation. 

Thus, we have established only the part of the united Dirac interaction of 
C-noninvariant character in which the introduced massless field $A_{\mu}^{A}$ is equalized 
with an axial-vector Coulomb field. Therefore, it is not surprising that the full Lagrangian 
$L^{A}$ invariant concerning the local gauge transformations has the following structure:
\begin{equation}
L^{A}=\overline{\psi}_{{l}^{A}}
\gamma_{5}(i\gamma_{\mu}\partial_{\mu}-m_{{l}^{A}})\psi_{{l}^{A}}-
\frac{1}{4}F_{\mu\lambda}^{A}F^{\mu\lambda}_{A}+
e_{l^{A}}\overline{\psi}_{{l}^{A}}\gamma_{5}\gamma_{\mu}\psi_{{l}^{A}}A_{\mu}^{A}.
\label{134}  
\end{equation}

The first pair of terms here corresponds to the free moving of a C-odd lepton of mass 
$m_{{l}^{A}},$ the term with a gauge-invariant tensor
\begin{equation}
F_{\mu\lambda}^{A}=\partial_{\mu}A_{\lambda}^{A}-\partial_{\lambda}A_{\mu}^{A}
\label{135}
\end{equation}
characterizes the free axial-vector Coulomb field, and the latter term describes the minimal interaction with this field of emission of an axial-vector current
\begin{equation}
j_{\mu}^{\gamma^{A}}=
\overline{\psi}_{{l}^{A}}\gamma_{5}\gamma_{\mu}\psi_{{l}^{A}}.
\label{136}  
\end{equation} 

Comparing (\ref{136}) with the anapole part of current (\ref{2}) at $g_{1l}=1,$ it is easy 
to observe the correspondence which says about that an axial-vector photon appears as the 
field $A_{\mu}^{A}$ necessary for the creation of a theory of truly neutral fermions 
invariant concerning the local axial-vector gauge transformations.

These transformations are the abelian ones. In other words, each of them remains as the 
transformation of an abelian group $U_{A}(1)$ if $\beta (x)$ in matrix $U_{A}$ is not 
connected with any so far unobserved properties of matter fields violating its unitarity.

Returning to (\ref{133}), we remark that the Dirac equation for a truly neutral lepton 
in the new field $A_{\mu}^{A}$ has the form
\begin{equation}
i\gamma_{\mu}\partial_{\mu}\psi_{{l}^{A}}=
m_{{l}^{A}}\psi_{{l}^{A}}-e_{l^{A}}\gamma_{\mu}A_{\mu}^{A}\psi_{{l}^{A}}.
\label{137}  
\end{equation}

From its point of view, the divergence of an axial-vector current is equal to
\begin{equation}
\partial_{\mu}j_{\mu}^{\gamma^{A}}=
2im_{{l}^{A}}\overline{\psi}_{{l}^{A}}\gamma_{5}\psi_{{l}^{A}}
\label{138}  
\end{equation} 
and becomes continuity equation only in the case of the lepton mass being wholly absent. 

At first sight in conformity with ideas of axial anomaly, this would have no place, since an 
ultimate expression [35-37] for the divergence of an axial-vector current $j_{\mu 5}$ in the conservation limit of a vector current $j_{\mu}$ is reduced to the following:
\begin{equation}
\partial_{\mu}j_{\mu 5}=
2im\overline{\psi}\gamma_{5}\psi
+\frac{e^{2}}{16\pi^{2}}\epsilon_{\mu\lambda\rho\sigma}F_{\mu\lambda}F_{\rho\sigma},
\label{139}  
\end{equation} 
where $e$ is the electron charge, $\epsilon_{0123}=1.$ 

On the other hand, the same lepton may not be simultaneously both a vector C-even fermion and 
an axial-vector C-odd one, as follows from considerations of symmetry. Therefore, we cannot 
exclude an axial-vector photon as the field giving in principle the possibility for the 
establishment in nature of continuity of an axial-vector current regardless of whether 
or not an unbroken vector current conservation law exists.

It is interesting, however, that the same photon of an axial-vector nature may not serve as the source of the two types of axial fields of a different character. At the same time, the existence itself of an electromagnetic field is explained by the fundamental symmetry between the electricity and the magnetism stating that to any type of the photon with the electric mass [38] 
and charge [39] corresponds a kind of monophoton [40] with the magnetic mass and charge. If such 
pairs are of axial-vector types, they constitute the naturally united axial-vector electromagnetic 
field. But a notion about axial-vector photons of a magnetic nature was introduced [41-44] 
over forty years ago. 

Another important prediction of a classification of elementary particles with respect 
to C-operation is the difference between the photons of a vector and an axial-vector
nature. From its point of view, the vector photons due to their C-invariant characters
have no interaction with truly neutral leptons but possess with leptons of C-even 
nature a kind of vector interaction.  

However, in spite of this, we can accept, in the framework of the standard electroweak 
theory, the vector photon $(\gamma^{V})$ as a gauge field if we make in (127)-(137) the 
following replacements:
\begin{equation}
A\rightarrow V, \, \, \, \, \gamma_{5}\rightarrow 1, \, \, \, \, 
\beta (x)\rightarrow \alpha (x), \, \, \, \, A_{\mu}^{A}\rightarrow A_{\mu}^{V}.
\label{140}  
\end{equation} 

Of them $\alpha (x)$ denotes the local phase in the transformation
\begin{equation}
\psi'_{{l}^{V}}(x)=U_{V}\psi_{{l}^{V}}(x), \, \, \, \, U_{V}=e^{i\alpha (x)}.
\label{141}  
\end{equation} 

Thus, only the part is obtained of the united Dirac interaction of a C-invariant nature 
in which a vector photon appears as the field $A_{\mu}^{V}$ necessary for the creation of 
a theory of Dirac fermions of C-even character invariant concerning the local vector gauge transformations of an unitary group $U_{V}(1)$ if $\alpha (x)$ in matrix $U_{V}$ does not 
possess any new properties violating its abelianity.

\vspace{0.8cm}
\noindent
{\bf 7 Unity of flavour and gauge symmetry laws}
\vspace{0.4cm}

It is already clear from the above reasoning that $U_{V}(1)$ and $U_{A}(1)$ describe the 
internal symmetry of C-invariant and C-noninvariant fermions, and an invariance of a theory concerning the local gauge transformations leads to the appearance of the corresponding fields 
of emission of photons of vector and axial-vector character. Their nature has been created so 
that to each type of C-even or C-odd lepton corresponds a kind of flavour. 

Furthermore, if these situations follow from a unified principle, flavour symmetry one 
must consider as a gauge symmetry [4].

To show their features, we investigate here those Lagrangians which may symbolically 
be written as
\begin{equation}
L_{0}=L_{0}^{V}+L_{0}^{A},
\label{142}  
\end{equation} 
\begin{equation}
L_{int}=L_{int}^{V}+L_{int}^{A}.
\label{143}  
\end{equation} 

At the same time, it is clear that (\ref{127}) and (\ref{131}) at the use of (\ref{140}) define $L_{0}^{V},$ $L_{int}^{V}$ and require the elucidation of the ideas of each of flavour and gauge symmetry laws.

From the point of view of a classification of elementary particles with respect to C-operation, 
any of (\ref{142}) and (\ref{143}) assumed that $U(1)\rightarrow U_{V}(1)$ or $U_{A}(1).$ In 
other words, among the fermions and photons there are vector and axial-vector particles. 
This in turn implies that
\begin{equation}
U_{A}\psi_{{l}^{V}}(x)=0, \, \, \, \, U_{A}\psi_{{l}^{A}}(x)\neq 0,
\label{144}  
\end{equation} 
\begin{equation}
U_{V}\psi_{{l}^{A}}(x)=0, \, \, \, \, U_{V}\psi_{{l}^{V}}(x)\neq 0.
\label{145}  
\end{equation} 

Under such circumstances, a full Lagrangian is invariant concerning the local gauge 
transformations: 
\begin{equation}
L_{0}'+L_{int}'=L_{0}+L_{int}.
\label{146}  
\end{equation} 

If we suppose here that 
$\psi_{{l}_{L,R}}=l_{L,R}$ and $\overline{\psi}_{l_{R,L}}=\overline{l}_{R,L},$
we would present the individual difermions (\ref{14}) and (\ref{20}) as follow:
\begin{equation}
(\psi_{l_{L,R}^{V}}, 
\overline{\psi}_{l_{R,L}^{V}}), \, \, \, \,
(\psi_{l_{L,R}^{A}}, \overline{\psi}_{l_{R,L}^{A}}),
\label{147}
\end{equation}
in which appears a fundamental part of flavour and gauge symmetries. 

In the framework of the standard electroweak theory, the same lepton must possess with  
vector and axial-vector fields of emission of the same types of photons simultaneously each 
of C-invariant and C-noninvariant types of interactions. They are of course described by the 
same Lagrangian that unites their in a unified whole. Therefore, from its point of view, it 
should be expected that $U(1)\rightarrow U_{V}(1)$ $\otimes$ $U_{A}(1).$ This standard 
unification leads us once more to an implication that
\begin{equation}
U_{A}\psi_{{l}^{V}}(x)\neq 0, \, \, \, \, U_{A}\psi_{{l}^{A}}(x)\neq 0,
\label{148}  
\end{equation} 
\begin{equation}
U_{V}\psi_{{l}^{A}}(x)\neq 0, \, \, \, \, U_{V}\psi_{{l}^{V}}(x)\neq 0.
\label{149}  
\end{equation} 

At these situations, (\ref{142}) and (\ref{143}) replace (\ref{146}) by
\begin{equation}
L_{0}'+L_{int}'=L_{0}+L_{int}+
e_{l^{V}}j_{\mu}A_{\mu}^{A}+e_{l^{A}}j_{\mu 5}A_{\mu}^{V},
\label{150}  
\end{equation} 
where it has been accepted that
\begin{equation}
e_{l^{A}}=e_{l^{V}}, \, \, \, \, A_{\mu}^{A}\neq A_{\mu}^{V},
\label{151}  
\end{equation}
\begin{equation}
j_{\mu}=\overline{\psi}_{{l}^{V}}\gamma_{\mu}\psi_{{l}^{V}}, \, \, \, \,
j_{\mu 5}=\overline{\psi}_{{l}^{A}}\gamma_{5}\gamma_{\mu}\psi_{{l}^{A}}.
\label{152}.
\end{equation} 

Each term with the current and the field of a different nature violates, as has been mentioned 
above, a gauge invariance. Their presence can explain, in the case of interaction [20] between 
the lepton and the field of emission, the appearance [12,13] of mixed difermions (\ref{47}), (\ref{48}) and the following connections:
\begin{equation}
(\psi_{l_{L}^{V}}, \overline{\psi}_{l_{R}^{A}}), \, \, \, \,
(\psi_{l_{R}^{V}}, \overline{\psi}_{l_{L}^{A}}),
\label{153}
\end{equation}
\begin{equation}
(\psi_{l_{L}^{A}}, \overline{\psi}_{l_{R}^{V}}), \, \, \, \,
(\psi_{l_{R}^{A}}, \overline{\psi}_{l_{L}^{V}}),
\label{154}
\end{equation}
which take place at the absence of mirror symmetry expressing the idea of flavour symmetry. 

This gives the right to interpret the flavour symmetry as a gauge invariance [4]. At the 
same time, gauge symmetry itself must be accepted as a mirror symmetry [45].

\vspace{0.8cm}
\noindent
{\bf 8 Conclusion}
\vspace{0.4cm}

Between the spaces $U_{V}(1)$ and $U_{A}(1)$ there exist fundamental differences, owing to 
which all elementary particles are united in families of a different C-parity. Thereby, they 
relate the flavour and gauge symmetries as a consequence of force unification forming the 
two left (right)-handed leptons in individual dileptons of the same vector or axial-vector 
character. Such a correspondence principle can explain the absence in nature of 
mixed diphotons 
\begin{equation}
(\gamma^{V}_{L}, \overline{\gamma}^{A}_{R}), \, \, \, \,
(\gamma^{V}_{R}, \overline{\gamma}^{A}_{L}),
\label{155}
\end{equation}
\begin{equation}
(\gamma^{A}_{L}, \overline{\gamma}^{V}_{R}), \, \, \, \,
(\gamma^{A}_{R}, \overline{\gamma}^{V}_{L})
\label{156}
\end{equation}
and the availability of a hard connection between the two left (right)-handed photons
in individual diphotons 
\begin{equation}
(\gamma^{V}_{L}, \overline{\gamma}^{V}_{R}), \, \, \, \,
(\gamma^{V}_{R}, \overline{\gamma}^{V}_{L}),
\label{157}
\end{equation}
\begin{equation}
(\gamma^{A}_{L}, \overline{\gamma}^{A}_{R}), \, \, \, \,
(\gamma^{A}_{R}, \overline{\gamma}^{A}_{L}).
\label{158}
\end{equation}

From this point of view, any of leptonic strings testifies in favor of the existence of a 
kind of bosonic string that unites the two left (right)-handed particles from the corresponding 
types of gauge bosons. 

Therefore, the diphotons (\ref{157}) and (\ref{158}) together with (\ref{147}) lead us 
to (\ref{89})-(\ref{91}) once more, confirming that the nature itself does not exclude 
the existence of both vector and axial-vector types of fermions.

To understand their unidenticality at a more concrete experimental level, one must refer to the inequality of the absolute values of the finding bounds [33] on $\mu_{l^{V}}=f_{2l^{V}}(0)$ and 
$d_{l^{A}}=g_{2l^{A}}(0),$ namely, on the sizes of the C-invariant magnetic and C-noninvariant electric dipole moments of leptons and other types of fermions, because they can appear, respectively, in the interaction vector and axial-vector structure dependence of particles 
and fields.

At the same time, it is clear that each type of gauge boson of a naturally united interaction constitutes a kind of current. This implies that not only the photon, but also the other 
boson leptonic currents include both vector C-even and axial-vector C-odd components.

In these circumstances, it seems possible to separate all gauge bosons into the two classes. 
The first of them consists of C-invariant vector bosons. They are of course the mediate bosons 
of vector types of interactions. We include in this class the vector photons $\gamma^{V}$ and 
weak $W^{\pm}$-bosons. A new example of the first group may be weak vector $Z^{\pm}$-bosons. 
To the second class apply the axial-vector C-noninvariant bosons. They come forward in the system 
as the mediate bosons of axial-vector types of interactions. A beautiful example is axial-vector photons $\gamma^{A}$ and weak $Z^{0} (W^{0})$-bosons. If we recognize such a behavior of the mediate bosons, accepting its ideas about that a classification of particle fundamental interactions and fields with respect to C-operation is fully compatible with gauge invariance, we would change our earlier presentations about the unified field theory of elementary particles. 

Finally, insofar as the neutrino oscillation is concerned, an old picture of its building
encounters the condition of a unity of flavour and gauge symmetry laws and requires the 
explanation both from the point of view of a classification of elementary particles with 
respect to C-operation and from the point of view of neutrino strings. To solve this question, 
it is desirable to recall at first the lepton and true flavours, the conservation of which unites the two left (right)-handed neutrinos in flavourless difermions of C-even or C-odd character. They have no definite mass. Therefore, in conformity with a gauge invariance principle, we conclude that the neutrino oscillations are carried out in the systems of the same vector or axial-vector types 
of neutrinos without flavour symmetry violation by the schemes
\begin{equation}
(\nu_{eL,R}, {\bar \nu_{eR,L}})\rightarrow 
(\nu_{\mu R,L}, {\bar \nu_{\mu L,R}}),
\label{159}
\end{equation}
\begin{equation}
(\nu_{\mu L,R}, {\bar \nu_{\mu R,L}})\rightarrow 
(\nu_{\tau R,L}, {\bar \nu_{\tau L,R}}),
\label{160}
\end{equation}
\begin{equation}
(\nu_{eL,R}, {\bar \nu_{eR,L}})\rightarrow
(\nu_{\tau R,L}, {\bar \nu_{\tau L,R}}).
\label{161}
\end{equation}

In the presence of a purely Coulomb part of current, they correspond to the interconversions 
of photons of the different components:
\begin{equation}
(\gamma_{L,R}, \overline{\gamma}_{R,L})\rightarrow
(\gamma_{R,L}, \overline{\gamma}_{L,R}).
\label{162}
\end{equation}

From such a point of view, the transitions (\ref{159})-(\ref{161}) become possible until an 
external action separates the photon strings (\ref{157}) and (\ref{158}) by a part in their structural particle type dependence. Therefore, to reanalyse the experiments about mixing angles 
at the new level, one must elucidate the ideas of each of those oscillations which originate, for example, in the systems of the quarks and other types of hadrons.

But at a given stage we can only add that regardless of the oscillating particle types, their 
mixing has the unified nature. Thereby, it reflects the unity of micro-world symmetry laws.

\vspace{0.8cm}
\noindent
{\bf References}
\begin{enumerate}
\item
B. Pontecorvo, B.: JETP {\bf 33}, 549 (1957)
\item
Fukuda, Y. et. al.: Phys. Rev. Lett. {\bf 81}, 1562 (1998)
\item
Wang, W.: AIP Conf. Proc. {\bf 1222}, 494 (2010). arXiv:0910.4605 [hep-ex]
\item
Sharafiddinov, R.S.: Bull. Am. Phys. Soc. {\bf 59(5)}, L1.00036 (2014)
\item
Rosenbluth, M.N.: Phys. Rev. {\bf 79}, 615 (1950)
\item
Lee, T.D., Yang, C.N.: Phys. Rev. {\bf 104}, 254 (1956)
\item
Zel'dovich, Ya.B.: JETP {\bf 33}, 1531 (1957)
\item
Zel'dovich, Ya.B., Perelomov, A.M.: JETP {\bf 39}, 1115 (1960)
\item
Landau, L.D.: JETP {\bf 32}, 405 (1957)
\item
Landau, L.D.: Nucl. Phys. {\bf 3}, 127 (1957)
\item
Sharafiddinov, R.S.: Bull. Am. Phys. Soc. {\bf 60(4)}, T1.00033 (2015). Spacetime 
Subst. {\bf 3}, 86 (2002). arXiv:physics/0305009
\item
Sharafiddinov, R.S.: In: Proceedings of the April Meeting of the American Physical 
Society, Jacksonville, Florida, April 14-17, 2007. Abstract, K11.00008
\item
Sharafiddinov, R.S.: J. Phys. Nat. Sci. {\bf 4}, 1 (2013). arXiv:physics/0702233
\item
Sharafiddinov, R.S.: In: Proceedings of the 75th Annual Meeting of the Southeastern 
Section of the American Physical Society, Raleigh, North Carolina, October 29 - November 1,
2008. Abstract, NB.00009
\item
Sharafiddinov, R.S.: Eur. Phys. J. Plus {\bf 126}, 40 (2011). arXiv:0802.3736 [physics.gen-ph]
\item
Sharafiddinov, R.S.: In: Proceedings of the April Meeting of the American Physical
Society, Dallax, Texas, April 22-25, 2006. Abstract, D1.00076
\item
Sharafiddinov, R.S.: Fizika {\bf B 16}, 1 (2007). arXiv:hep-ph/0512346
\item
Sharafiddinov,R.S.: Spacetime Subst. {\bf 3}, 134 (2002). arXiv:physics/0305015
\item
Yuldashev, B.S., Sharafiddinov, R.S.: Spacetime Subst. {\bf 5}, 137 (2004). 

arXiv:hep-ph/0510080
\item
Begzhanov, R.B., Sharafiddinov, R.S.: Mod. Phys. Lett. {\bf A 15}, 557 (2000)
\item
Greenberg, O.W.: Phys. Rev. Lett. {\bf 89}, 231602 (2002). arXiv:hep-ph/0201258
\item
Colladay, D., Kostelecky, A.: Phys. Rev. {\bf D 58,} 116002 (1998). arXiv:hep-ph/9809521
\item
Glashow, S.L.: Nucl. Phys. {\bf 22}, 579 (1961)
\item
Salam, A., Ward, J.C.: Phys. Lett. {\bf 13}, 168 (1964)
\item
Weinberg, S.: Phys. Rev. Lett. {\bf 19}, 1264 (1967)
\item
Chaichian, M., Fujikawa, K., Tureanu, A.: arXiv:1103.0168 [hep-th]
\item
Sharafiddinov, R.S.: Phys. Essays {\bf 19}, 58 (2006). arXiv:hep-ph/0407262
\item
Majorana, E.: Nuovo Cimento {\bf 14}, 171 (1937)
\item 
Sharafiddinov, R.S.: Bull. Am. Phys. Soc. {\bf 59(5)}, T1.00004 (2014)
\item
Zel'dovich, Ya.B.: Dokl. Akad. Nauk SSSR {\bf 91}, 1317 (1953)
\item
Konopinsky, E.J., Mahmoud, H.: Phys. Rev. {\bf 92}, 1045 (1953)
\item
Sharafiddinov, R.S.: In: Proceedings of the 2rd Eurasian Conference on Nuclear Science 
and Its Application, Almaty, September 16-19, 2002 (Almaty, Kazakhstan, 2002), 
Abstracts, p. 146.
\item
Yao, W.-M. et al.: Particle Data Group. J. Phys. {\bf G 33}, 1 (2006)
\item
Zel'dovich, Ya.B.: Dokl. Akad. Nauk SSSR {\bf 86}, 505 (1952)
\item
Adler, S.L.: Phys. Rev. {\bf 177}, 2426 (1969)
\item
Adler, S.L., Bardeen, W.A.: Phys. Rev. {\bf 182}, 1517 (1969)
\item
Bell, J., Jackiw, R.: Nuovo Cimento {\bf 51}, 47 (1969)
\item
Prokopec, T., Tornkvist, O., Woodard, R.: Phys. Rev. Lett. {\bf 89}, 101301 (2002).

arXiv:astro-ph/0205331
\item
Altschul, B.: Phys. Rev. Lett. {\bf 98}, 261801 (2007). arXiv:hep-ph/0703126 
\item
Sharafiddinov, R.S.: Bull. Am. Phys. Soc. {\bf 59(5)}, T1.00005 (2014). Spacetime 
Subst. {\bf 3}, 132 (2002). arXiv:physics/0305014
\item
Salam, A.: Phys. Lett. {\bf 22}, 683 (1966)
\item
Taylor, J.D.: Phys. Rev. Lett. {\bf 18}, 713 (1967)
\item
Markov, M.A.: Zh. Eksp. Teor. Fiz. {\bf 3}, 98 (1966)
\item
Slad', L.M.: Yad. Fiz. {\bf 27}, 1417 (1978)
\item
Giveon, A., Witten, E.: Phys. Lett. {\bf B 332}, 44 (1994). arXiv:hep-th/9404184
\end{enumerate}
\end{document}